\documentclass[preprint,showpacs,preprintnumbers,amsmath,amssymb,
superscriptaddress,unsortedaddress]{revtex4}

\usepackage{graphicx}% Include figure files
\usepackage{dcolumn}% Align table columns on decimal point
\usepackage{bm}% bold math

\begin{document}

%\preprint{APS/123-QED}

\title{Examples of nonuniform limiting distributions
for the quantum walk on even cycles}

\author{Ma{\l}gorzata Bednarska}
%\email{mbed@amu.edu.pl}
\affiliation{Faculty of Mathematics
and Computer Science,
 Adam Mickiewicz University,
Umultowska 87, 61-614 Pozna\'{n}, Poland.
}
\author{Andrzej Grudka}%
%\email{agie@amu.edu.pl}
\affiliation{Faculty of Physics,
 Adam Mickiewicz University,
Umultowska 85, 61-614 Pozna\'{n}, Poland.
}
\author{Pawe{\l} Kurzy\'nski}
%\email{@amu.edu.pl}
\affiliation{Faculty of Physics,
 Adam Mickiewicz University,
Umultowska 85, 61-614 Pozna\'{n}, Poland.
}
\author{Tomasz {\L}uczak}
%\email{tomasz@amu.edu.pl}
\affiliation{Faculty of Mathematics
and Computer Science,
 Adam Mickiewicz University,
Umultowska 87, 61-614 Pozna\'{n}, Poland.
}
\author{Antoni W\'ojcik}
%\email{antwoj@amu.edu.pl}
\affiliation{Faculty of Physics,
 Adam Mickiewicz University,
Umultowska 85, 61-614 Pozna\'{n}, Poland.
}

\begin{abstract}
In the note we show how the choice of the initial states
can influence the evolution of time-averaged
probability distribution of the quantum walk on even
cycles.
\end{abstract}

\maketitle

The analysis of discrete quantum random walks initiated by
Aharonov {\sl et al.} \cite{Aharonov} and its possible
applications for constructing efficient  quantum algorithms
(\cite{Shenvi}-\cite{Amb3}) has recently attracted a lot of
attention. Although many questions in this area  remain open, it
is well known that the behaviour  of classical and quantum walks
can be very different, as it can be seen by studying  spreading,
mixing and hitting times (\cite{Aharonov}, \cite{Amb} and
\cite{Kempe2}) or limiting distributions \cite{Bed}. One of the
differences between quantum and classical walks we explore in this
note is that one can start a quantum walk not from a single
occupied node, but from the superposition of many nodes. The
influence of the initial state on the behaviour of  a quantum walk
was studied by Tragenna {\sl et al.} \cite{Tragenna}.
In~\cite{Bed} we mentioned the possibility of generating highly
nonuniform limiting distributions in a quantum walk on even cycles
starting from superposition states. In this note we show that the
initial conditions can affect  the time evolution of the total
variation distance of time-averaged probability distribution in a
decisive way.

We shall study  a quantum random walk on an even cycle with
$d$ nodes, using a  model proposed by Aharonov
{\sl et al.}~\cite{Aharonov}. In this setting the
nodes of the cycle
are represented by vectors $|v\rangle$, $v=0,1,\dots,d-1$,
which form an orthonormal basis of the Hilbert space $H_V$. An
auxiliary two-dimensional Hilbert space $H_A$ (coin space) is
spanned by vectors $|s\rangle$, $s=0,1$. The initial state of the
walk is a normalized vector
\begin{equation}\label{eq0}
|\Psi_0\rangle=\sum_{s,v}\gamma_{sv}|s,v\rangle=\sum_{s,v}\gamma_{sv}|s\rangle|v\rangle
\end{equation}
from the  tensor product $H=H_A\otimes H_V$. In a single step of the walk the
state changes according to the equation
\begin{equation}\label{eq1}
|\Psi_{t+1}\rangle=U|\Psi_t\rangle,
\end{equation}
where the operation $U=S(H\otimes I)$
first applies the Hadamard gate operator
$H=\frac{1}{\sqrt{2}}\sum_{s,s'}(-1)^{ss'}|s\rangle\langle s '|$
to the vector from $H_A$, and then
shifts the state  by the operator
\begin{equation}\label{eq2}
S= \sum_{s,v}|s\rangle\langle s |\otimes|v+2s-1 ({\rm mod}\  d)\rangle\langle v|\,.
\end{equation}

The operator $U$ has been studied in \cite{Bed}, where we prove that
\begin{equation}\label{eigenval}
U |\phi_{jk}\rangle=c_{jk}|\phi_{jk}\rangle,
\end{equation}
where the eigenvalues $c_{jk}$ are given by
\begin{equation}\label{eq7}
c_{jk}=\frac{1}{\sqrt{2}}\big( (-1)^k\sqrt{1+\cos^2(2\pi j/d)}-
i\sin(2\pi j/d) \big),
\end{equation}
for $k=0,1$, and $j=0,1,\dots,d-1$, while the corresponding
eigenvectors are
\begin{equation}\label{eq8}
|\phi_{jk}\rangle=(a_{jk}|0\rangle+a_{jk}b_{jk}|1\rangle)\otimes\sum_v\omega_d^{jv}|v\rangle,
\end{equation}
where $\omega_d=e^{2\pi i/d}$,
\begin{align}
a_{jk}&=1\Big/\sqrt{d(1+|b_{jk}|^2)},\label{eq9}\\
b_{jk}&=\omega_d^j\big( (-1)^k\sqrt{1+\cos^2(2\pi j/d)}-
\cos(2\pi j/d) \big).\label{eq10}
\end{align}
The probability distribution on the nodes of the cycle after the
first $t$ steps of the walk is given by
\begin{equation}\label{eq3}
p_t(v)=\sum_s|\langle s,v|\Psi_t\rangle|^2.
\end{equation}

As was observed by Aharonov {\it  at el.\/}~\cite{Aharonov}, for a
fixed $v$, the probability  $p_t(v)$ is `quasi-periodic' as a
function of $t$ and thus, typically, it does not converge to a
limit. Thus, instead of $p_t(v)$, the authors of~\cite{Aharonov}
propose to consider time-averaged probability distribution
\begin{equation}\label{eq33}
\bar p_t(v)=\frac{1}{t+1}\sum_{i=0}^t p_t(v),
\end{equation}
and its limiting distribution
\begin{equation}\label{eq34}
\pi(v)=\lim_{t\to\infty}\bar  p_t(v).
\end{equation}
In order to present the global properties of the walk let us also
define the total variation distance
\begin{equation}
  \label{eqTVD}
  \Delta_t=\frac{1}{2}\sum_{v=0}^{d-1}\Big|\bar p_t(v)-\frac{1}{d}\Big|,
\end{equation}
which measures how far is time - averaged probability distribution
from uniform distribution. $\Delta_t$ tends to limit which will be
denoted by $\Delta_{\infty}=\lim_{t\to\infty}\Delta_t$.

%%%%%%%%%%%%%%%%%%%%%%%%%%%%%%%%%

\begin{figure}[h] %% h=here, t=top,p=page of floats
\centerline{\resizebox{0.5\textheight}{!}{\rotatebox{270}
{\includegraphics{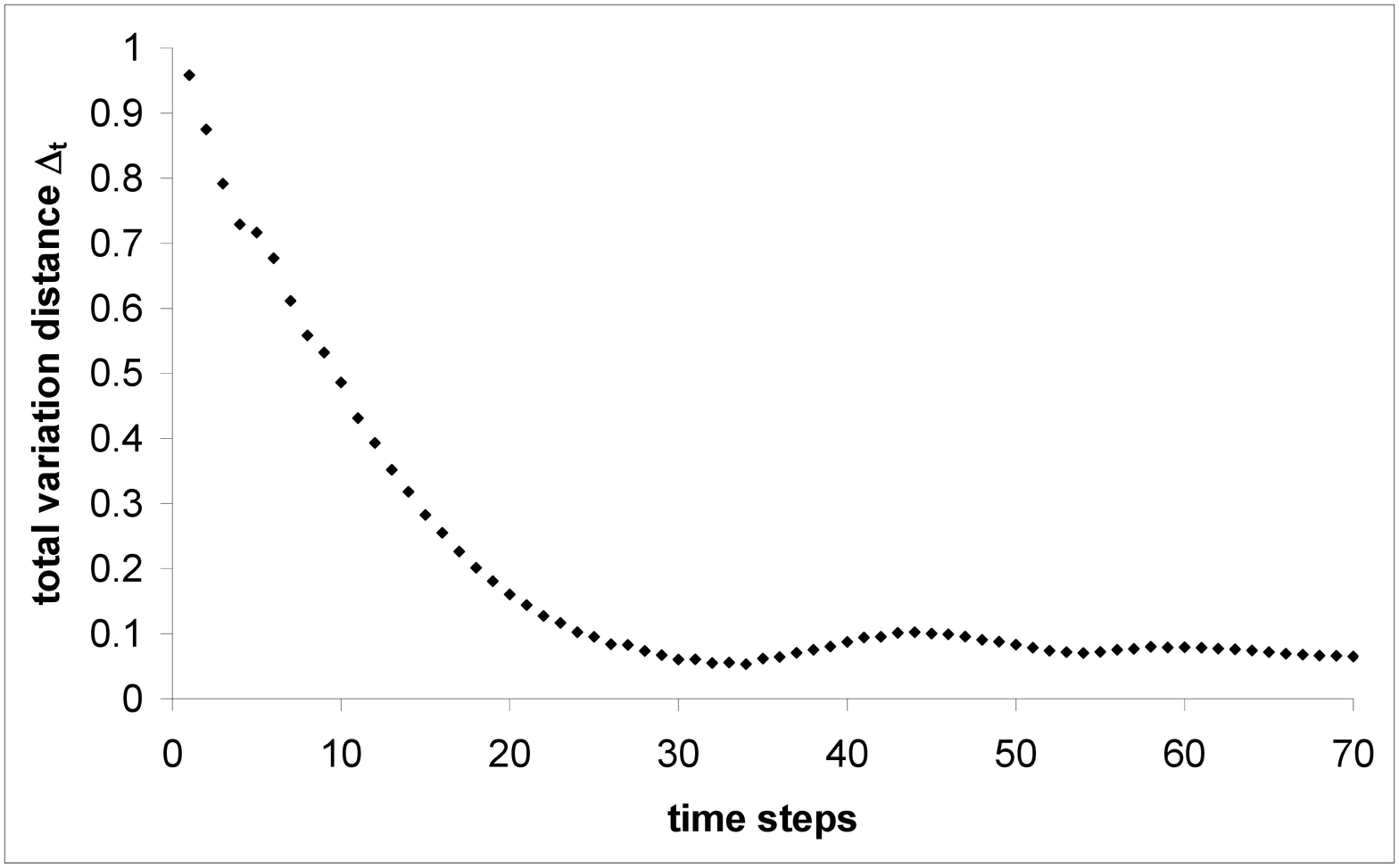}}}} \caption{Time evolution of the total
variation distance $\Delta_t$ for the initial state
$|\Psi_{0}^{(2d)}\rangle$ ($d=24$).}
 \label{g1}
\end{figure}

%\begin{figure}{\rotatebox{270}{
%\includegraphics[height=3cm] {fig1.eps}}}  %Here is how to import EPS art
%\caption{\label{g1}Time evolution of the total variation distance
%$\Delta_n$ for the initial state $|\Psi_{0}^{(2d)}\rangle$.}
%\end{figure}

\begin{figure}[h] %% h=here, t=top,p=page of floats
\centerline{\resizebox{0.5\textheight}{!}{\rotatebox{270}
{\includegraphics{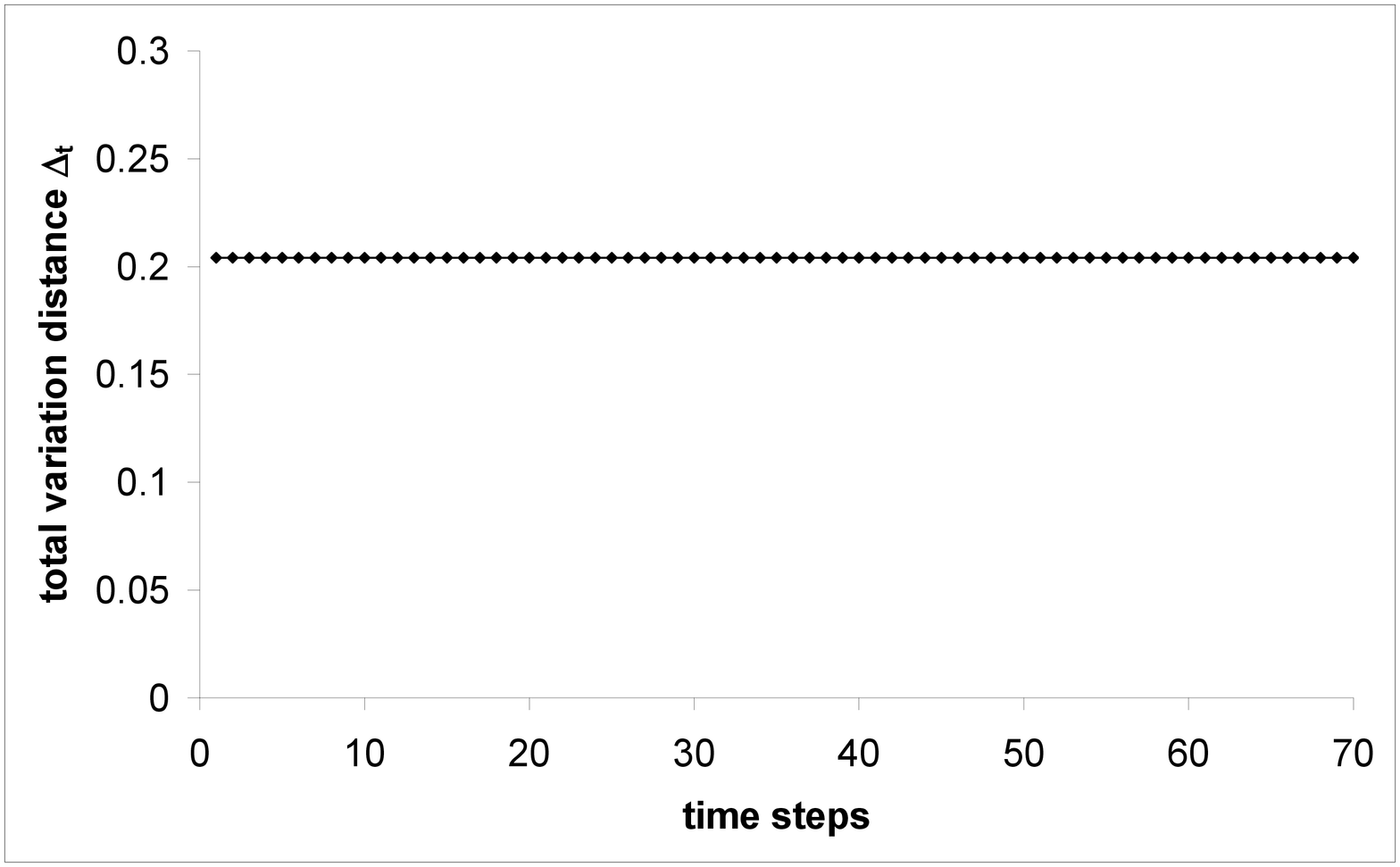}}}}%Here is how to import EPS art
\caption{\label{g2} Time evolution of the total variation distance
$\Delta_t$ for the initial state $|\Psi_{0}^{(2)}\rangle$
($d=24$). Diamonds -- numerical simulations, line -- analytical
value of Eq. (\ref{eqs7}).}
\end{figure}

\begin{figure}[h] %% h=here, t=top,p=page of floats
\centerline{\resizebox{0.5\textheight}{!}{\rotatebox{270}
{\includegraphics{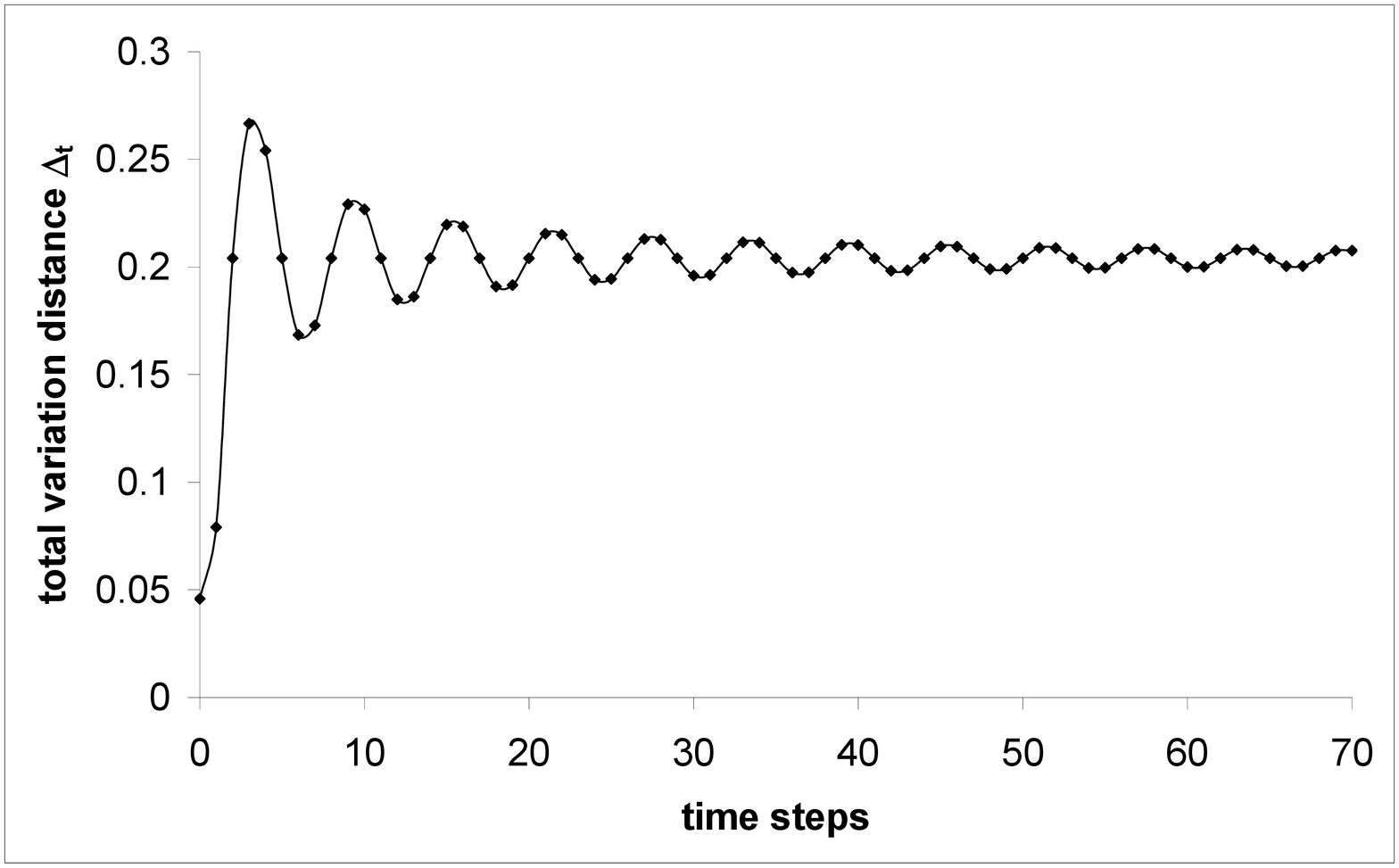}}}}% Here is how to import EPS art
\caption{\label{g3} Time evolution of the total variation distance
$\Delta_t$ for the initial state $|\Psi_{0}^{(4)}\rangle$
($d=24$). Diamonds -- numerical simulations, line -- analytical
value of Eqs. (\ref{eqTVD}) and (\ref{eqs10}).}
\end{figure}

Figs.~\ref{g1}, \ref{g2}, and \ref{g3}, show  the evolution of the total
variation distance $\Delta_t$  for the case of three
different initial states: $|\Psi_{0}^{(2d)}\rangle$,
$|\Psi_{0}^{(2)}\rangle$ and $|\Psi_{0}^{(4)}\rangle$, which
can be written as a superposition of some $2d$, $2$, and $4$, eigenvectors
$|\phi_{jk}\rangle$, respectively.
Thus,
$|\Psi_{0}^{(2d)}\rangle=\sum_{jk}g_{jk}\omega_{d}^{-v_{0}j}|\phi_{jk}\rangle$
is the state with a single occupied node $v_{0}$ where
$g_{jk}=a_{jk}(1+ib_{jk}^{\ast})/\sqrt{2}$;
$|\Psi_{0}^{(2)}\rangle=\frac{1}{\sqrt{2}}\left(
|\phi_{3,0}\rangle + |\phi_{9,0}\rangle \right)$ is a
superposition of two degenerate eigenvectors; finally
$|\Psi_{0}^{(4)}\rangle=\frac{1}{\sqrt{2}}\left(
|\phi_{3,0}\rangle + |\phi_{9,0}\rangle - |\phi_{15,0}\rangle -
|\phi_{21,0}\rangle \right)$.

In the case a quantum walk starts with $|\Psi_{0}^{(2d)}\rangle$,
one observe decaying of the total variation distance to the nonzero
value $\Delta_{\infty}^{(2d)}$  (Fig.~\ref{g1}).
An analytic form of $\Delta_{\infty}^{(2d)}$ can be found in a similar
way as in \cite{Bed}  (we remark
that there is a minor error in the equation (22) in \cite{Bed}).
Thus, we get
\begin{equation}
    \label{eqp}
\pi (v)= \frac{1+f(s)-(-1)^{\xi}f(s')}{d},
\end{equation}
where
\begin{equation}
    \label{eqp2}
f(x)= \frac{\sqrt{2}}{1-(-z)^{d/2}}z^{x}-\delta_{x0}-\frac{1}{d},
\end{equation}
\begin{equation}
    \label{eqxi}
\xi= \frac{(1+(-1)^{d/2})}{2},
\end{equation}
(i.e., $\xi=1$ when $d/2$ is even, and $\xi=0$ if $d/2$ is odd), and
\begin{equation}
    \label{eqss}
s = s(v)=\min\left( |v-v_{0}|,d-|v-v_{0}| \right),
\end{equation}
denotes the distance between nodes $v_{0}$ and $v$, and
$s'=d/2-s$. If $d\gg1$ we can write $\Delta_{\infty}^{2d}$ in a
simple form. When $\xi=0$
\begin{equation}
    \label{eqs1}
\Delta_{\infty}^{(2d)}=1/d,
\end{equation}
while in the case of $\xi=1$
\begin{equation}
    \label{eqs2}
\Delta_{\infty}^{(2d)}=\frac{2}{d}-\frac{4}{d^{2}}\left(1-2\frac{\log_{2}d-1/2}{\log_{2}z}\right),
\end{equation}
where $z=3-2\sqrt{2}$. For the particular case presented at Fig.~\ref{g1},
(\ref{eqs2}) gives the value $\Delta_{\infty}^{(2d)}=0.054$. Hence,
if we start with a  single occupied node, the total variation
distance decreases steadily in time and its limiting value tends to zero
as the graph size $d$ grows. Let us present now two examples of
walk for which the dynamics of the total variation distance is
dramatically different. Fig.~\ref{g2} pictures the evolution of
a walk where  $\Delta_{t}^{(2)}\neq 0$ does not change in time.
It starts with the initial state  of the form
the form
\begin{equation}
    \label{eqs3}
\frac{1}{\sqrt{2}}\left(|\phi_{m,k}\rangle+|\phi_{d/2-m,k}\rangle\right),
\end{equation}
for $m=3$ and $k=0$. The state described by (\ref{eqs3}) for
$m=0,\dots,m_{\max}$ as well as the state
\begin{equation}
    \label{eqs4}
\frac{1}{\sqrt{2}}\left(|\phi_{d/2+m,k}\rangle+|\phi_{d-m,k}\rangle\right),
\end{equation}
for $m=1,\dots,m_{\max}$, ($m_{\max}=\lfloor(d-2)/4\rfloor$) consists
of two degenerated eigenvectors. Since the evolution of the
superposition of any number of degenerated eigenvectors leads only
to the global phase changes so the dynamics of the probability
distribution is frozen and $\pi (v) = \bar p_{t} (v)=p_{0} (v)$.
For the states given by (\ref{eqs3}) and (\ref{eqs4}) the limiting
distribution takes form
\begin{equation}
    \label{eqs5}
\pi (v) = \frac{1}{d} +
\frac{(-1)^{v}\sin\alpha}{d\sqrt{1+\cos^{2}\alpha}}\sin\left(\alpha\left(2v+1\right)\right),
\end{equation}
where $\alpha=2\pi m/d$. Thus
\begin{equation}
    \label{eqs6}
\Delta_{t}^{(2)} =
\frac{\sin\alpha}{d\sqrt{1+\cos^{2}\alpha}}\sum_{v}\big|
\sin\left(\alpha\left(2v+1\right)\right)\big|.
\end{equation}
When $m$ divides $d/2$ the summation can be easily perform leading
to
\begin{equation}
    \label{eqs7}
\Delta_{t}^{(2)} =
 \frac{m}{d}
 \frac{1}{\sqrt{1+\cos^{2}\alpha}}\left(1-\cos\left(2\alpha\left(\eta+1\right)\right)\right),
\end{equation}
where
\begin{equation}
    \label{eqs8}
\eta=\lfloor \frac{d}{4m}-\frac{1}{2} \rfloor.
\end{equation}
For $m=3$ and $d=24$ (\ref{eqs7}) gives $0.204$.

The last example of the time evolution, depicted at   Fig.~\ref{g3},
is, perhaps, most intriguing. The changes of the total variation distance
in this case resembles the motion of the damped harmonic
oscillator with shifted equilibrium. Let us emphasize also that
the limiting value of the total variation distance
$\Delta_{\infty}^{(4)}$ is much higher than the initial one
$\Delta_{0}^{(4)}$. The initial state $|\phi_{0}^{(4)}\rangle$ is
of the kind
\begin{equation}
    \label{eqs9}
\frac{1}{2}\left(|\phi_{m,k}\rangle+|\phi_{d/2-m,k}\rangle-|\phi_{d/2+m,k,k}\rangle-|\phi_{d-m,k}\rangle\right).
\end{equation}
The time-averaged probability
distribution for the initial states of the form given by
(\ref{eqs9}) can be described as
\begin{equation}
    \label{eqs10}
\bar p_{t} (v) = A(v) + B(v)\frac{\sin \left( 2\varphi_{mk}(t+1)
\right)}{t+1},
\end{equation}
where $A(v)=\pi (v)$, $B(v)=\left(p_{0}(v) -
\pi(v)\right)/\sin(2\varphi_{mk})$, $\varphi_{mk}$ is the phase of
the eigenvalue $c_{mk}$ $\left( c_{mk}=e^{i\varphi_{mk}} \right)$
and $\pi (v)$ is given by (\ref{eqs5}). Fig.~\ref{g3} presents
$\Delta_{t}^{(4)}$ calculated with the use of (\ref{eqTVD}) and
(\ref{eqs10}) as well as the results of numerical calculation.

In conclusion, we demonstrated how the initial conditions affects the
dynamics of the quantum walk on cycle. We gave examples
for three different kinds of behavior of the total variation
distance between given distribution and uniform distribution:
decaying, constant and damped oscillating.

%\vskip1truecm

\begin{acknowledgments}
A.G.\ and A.W.\ were supported by
 the State Committee for Scientific Research (KBN)
 grant 0~T00A~003~23. P.K. would like to
thank Adam Mickiewicz University and National University of
Singapore for support.
\end{acknowledgments}

%\bibliography{bezfaz}% Produces the bibliography via BibTeX.

\end{document}